\documentclass[aip,rsi,amsmath,amssymb,preprint]{revtex4-1}

\usepackage{amssymb,amsfonts,amsmath,graphicx}
\usepackage{subfigure}
\usepackage{color}
\usepackage{natbib}
\usepackage{algorithm}
\usepackage{algpseudocode}
\usepackage{algorithmicx }
\usepackage{amsthm}

\newcommand{\argmax}{\arg\!\max}

\begin{document}
\title{Information Theoretical Noninvasive Damage Detection In Bridge Structures}

\author{Amila Sudu Ambegedara}
\affiliation{Department of Mathematics, Clarkson University, 8 Clarkson Ave, Potsdam, New York, 13699-5815, USA} 

\author{Jie Sun}
\email{sunj@clarkson.edu}
\affiliation{Department of Mathematics, Clarkson University, 8 Clarkson Ave, Potsdam, New York, 13699-5815, USA} 
\affiliation{Department of Computer Science, Clarkson University, 8 Clarkson Ave, Potsdam, New York, 13699-5815, USA} 
\affiliation{Department of Physics, Clarkson University, 8 Clarkson Ave, Potsdam, New York, 13699-5820, USA} 
\author{Kerop Janoyan}
\affiliation{Department of Civil \& Environmental Engineering, Clarkson University, Potsdam, New York, 13699-5710, USA}

\author{Erik Bollt}
\email{bolltem@clarkson.edu}
\affiliation{Department of Mathematics, Clarkson University, 8 Clarkson Ave, Potsdam, New York, 13699-5815, USA}

\begin{abstract}
Damage detection of mechanical structures such as bridges is an important research problem in civil engineering. Using spatially distributed sensor time series data collected from a recent experiment on a local bridge in upper state New York, we study noninvasive damage detection using information-theoretical methods. Several findings are in order. First, the time series data, which represent accelerations measured at the sensors, more closely follow Laplace distribution than normal distribution, allowing us to develop parameter estimators for various information-theoretic measures such as entropy and mutual information. Secondly, as damage is introduced by the removal of bolts of the first diaphragm connection, the interaction between spatially nearby sensors as measured by mutual information become weaker, suggesting that the bridge is ``loosened". Finally, using a proposed oMII procedure to prune away indirect interactions, we found that the primary direction of interaction or influence aligns with the traffic direction on the bridge even after damaging the bridge.

\end{abstract}

\today

\maketitle

\smallskip
\noindent\textbf{Keywords.} {\small Noninvasive Damage Detection, Optimal Mutual Information, Bridge Structure}

\begin{quotation}
Bridges are important and overwhelmingly common for daily commuting and transportation. Because of their prevalence, it is important to monitor the mechanical ``health" of bridges especially to assess the risk of structural fatigue and more importantly to suggest timely maintenance in order to avoid sudden and disastrous collapse. Among the various techniques for structural health monitoring of bridges, we here consider information-theoretic measures which require minimal assumptions regarding the specific location, material, and age of the bridge. The data we use are time series collected on spatially distributed sensors from a controlled damage experiment performed on a local bridge in upper state New York. We found that the primary direct interactions underlie the bridge aligns with the traffic direction, and the bridge became effectively ``loosened" after introduction of the damage.

\end{quotation}

\section{Introduction.}

Damage detection of civil infrastructure such as bridges has gained considerable interest, for obvious economic and public safety reasons. Damage here can be described  as a change of material or geometrical properties that impact the performance of engineering systems  \cite{farrar2007introduction}. In the literature there are many traditional methods to detect the damage of a bridge\cite{wahab1999damage,kim2003damage,alvandi2006assessment,cruz2009performance}. Among them, non-invasive techniques are appropriate for many situations as they are non-destructive and often less expensive at least as a precursory scanning approach, in case the more expensive and direct inspection methods are prescribed.

As dynamical properties of healthy and damaged bridges differ, parameters such as natural frequencies, damping ratio and mode shapes can be used to detect the presence of damage in a bridge. The modal curvature method-vibration based damage identification technique, has been used extensively in literature as the stiffness of the structure directly relate to the natural frequencies and mode shapes, 
\cite{wahab1999damage,kim2003damage,alvandi2006assessment,cruz2009performance,sampaio1999damage,dutta2004damage}.

Model-based methods such as neural networks and genetic algorithms \cite{lee2005neural,mehrjoo2008damage,barai1995performance} have also been used as a basis to develop damage detection of bridges utilizing artificial intelligence and machine learning techniques. These methods have been be used to recognize patterns of the damaged and non-damaged systems. In \cite{lee2005neural}  the study used a genetic algorithm based damage detection method, wherein the authors formulated the structure damage as an optimization problem. Moreover they used static displacements as measured responses. In  \cite{barai1997time}, several drawbacks were discussed when adopting the traditional neural networks in dealing with patterns that vary over time. Also in  \cite{barai1997time}, time-delay neural networks were proposed to detect the damage of railway bridges and compared with traditional neural networks. 

In this paper, we propose a mutual information (MI) based damage detection of a highway bridge, where the specific pairwise interactions are analyzed successively by an interaction analysis to  uncover and detect interactions, and most importantly, changes in the way various regions of the structure may interact with each other as the system  becomes damaged. We develop what we call, optimal mutual information interaction (oMII) as explained in section \ref{sec:InfoMI} as analogous to our previously developed optimal causation entropy (oCSE) \cite{sun2014,sun2014entropy,sun2015causal}  as this is a method of uncovering {\it direct} interactions that are more indicatively sensitive  to  system changes. In the literature \cite{fraser1986independent,wells1996multi,battiti1994using,pluim2003mutual,jeong2001mutual,martinerie1992mutual,butte2000mutual,qiu2009fast} many researchers used MI based techniques to analyze different dynamical fields.

Since the working premise is that a damaged bridge's dynamics are different, and therefore so is the way vibrational energy may transmit through the structure, the MI of sensed accelerometry within the structure differs when comparing  healthy bridge's pairwise MI to that of the same bridge later measured in a damaged state. Therefore, we assert definitively a principle that damaged verses undamaged bridges have different dynamics. Specifically direct influences may change due to damage. So comparing influences as inferred by oMII over time we claim that we can non-invasively detect important (damage) changes. Here we consider a specific bridge as our test platform. The New York State Route 345 bridge crosses over Big Sucker Brook, in the town of Waddington, NY and it was constructed in 1957. The instrumentation has involved 30 dual-axial accelerometers placed on the bridge at 30 locations therein. The test protocol has involved collection of data from three levels of damage as introduced by removing bolts from a diaphragm \cite{whelan2009service}.

This paper organized as follows: We start the paper by introducing mathematical tools of information theory. 
The methods that we use to study the damage detection is given in the Secs,~\ref{subsec:PMI} and~\ref{subsec:method}. Section \ref{sec:damage} describes experimental protocol  of the bridge on which we study, and  damage  in controlled experiments, together with the experimental protocol and the nature of the instrumentation based on accelerometry sensing. The mathematical analysis reveals that the data sets more closely follow Laplace distributions than Gaussian distributions, and the analysis of this  can be found in section  \ref{subsec: Laplace}. Finally, in section  \ref{sec:results} we present the results that we obtained from pairwise mutual information and oMII, that we use to reveal notable changes between the damaged and undamaged bridge.

\section{Information-theoretic Measures.}\label{sec:InfoMI}

\subsection{Basics from Information Theory}

Since our analysis is based on developing the oMII which is an information theoretic quantity, first we review some basic background theory. 
In this section we review basic information-theoretic measures which will be useful in the application of damage detection, leading to the mutual information and conditional mutual information whose application will be described in section III and IV, including details of the relevant conditions.
Consider a continuous random variable $X$ whose probability density function is denoted by $p(x)$. Then the (differential) entropy of $X$, denoted by $h(X)$, is defined as~\cite{shannon1948,cover2006}
\begin{equation}\label{dist}
   h(X) =-\int p(x) \log p(x)dx.
\end{equation}
The entropy captures the uncertainty associated with a random variable.
Such definition naturally extends to the case of multiple (or multivariate) random variables. For example, the joint entropy of two random variables $X$ and $Y$ is given by~\cite{shannon1948,cover2006}
\begin{equation}\label{dist}
   h(X,Y) =-\int p(x,y) \log p(x,y)dxdy,
\end{equation}
where $p(x,y)$ is the joint density of $(X,Y)$. When conditional densities are involved, the corresponding entropy measures are called conditional entropies. For two random variables, there are two such conditional entropies~\cite{shannon1948,cover2006}:
\begin{equation}
\begin{cases}
h(X|Y) =-\int p(x,y) \log p(x|y)dxdy,\\
h(Y|X) =-\int p(x,y) \log p(y|x)dxdy,
\end{cases}
\end{equation}
where $p(x|y)$ and $p(y|x)$ are conditional densities.

\begin{figure}[htbp]
\subfigure[~Venn diagram for two variables]{\includegraphics[width=0.47\textwidth]{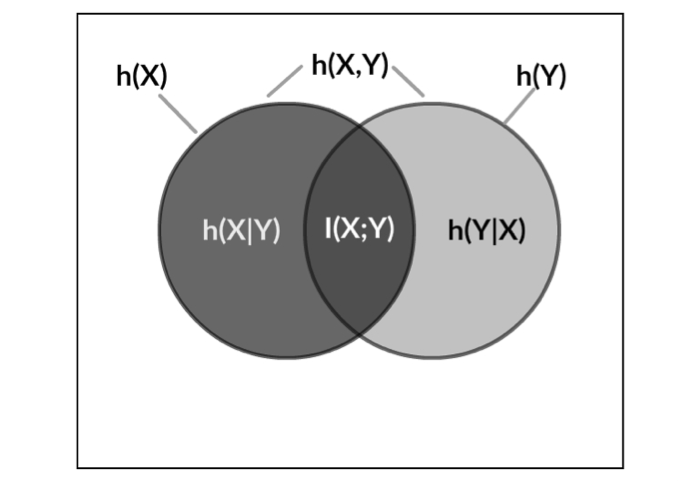}}
\subfigure[~Venn diagram for three variables]{\includegraphics[width=0.50\textwidth]{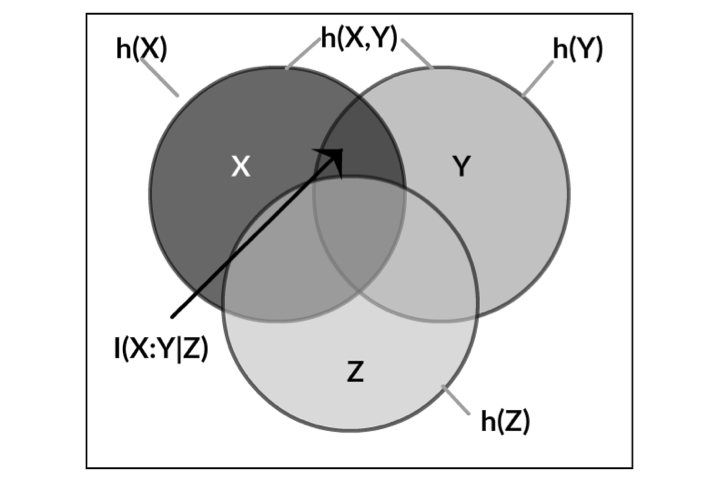}}
\caption{Information Venn diagram for two (a) and three (b) variables, respectively.}
\label{fig:mutualvenn}
\end{figure}

The joint and conditional entropies can be used to construct measures that detect the statistical dependence or independence between random variables as the case may be. For example, the mutual information (MI), given below~\cite{shannon1948,cover2006},
\begin{equation} \label{MI}
I(X;Y) = h(X) + h(Y) - h(X,Y),
\end{equation}
is nonnegative [$I(X;Y)\geq0$] and equals zero if and only if $p(x,y)=p(x)p(y)$, that is, $X$ and $Y$ are statistically independent.
It is often convenient to visualize the relationship among various entropies and mutual information in an information Venn diagram as shown in Fig.~\ref{fig:mutualvenn}(a).
When there is a third variable involved, the conditional mutual information (CMI) between $X$ and $Y$ given $Z$ is defined as follows~\cite{shannon1948,cover2006},
\begin{equation}\label{condMI}
	I(X;Y|Z) = h(X|Z) + h(Y|Z) - h(X,Y|Z).
\end{equation}
Like mutual information, the conditional mutual information obeys an analogous inequality, $I(X;Y|Z)\geq 0$, with equality if and only if $X$ and $Y$ are  independent given $Z$. The relation between various entropies, mutual information and conditional mutual information are visualized using an information Venn diagram in Fig.~\ref{fig:mutualvenn}(b). 
\vspace{1cm}

\subsection{Spatial Pairwise Mutual Information} \label{subsec:PMI}
In many engineering applications such as the monitoring of mechanical structures, sensors are often placed spatially. This motivates a concept of spatial pairwise mutual information. 

The idea of damage detection by information flow is to compare how signals measured from different spatial locations on the bridge respond to the challenge of the truck passage. Therefore, the hypothesis is that the signals compared between sites, pairwise, may show a given coincidence as measured by MI when the bridge is in a healthy state, but the transmission of vibrational energy between sites becomes different if the bridge has been altered or damaged. \textit{We assert that just as energy is transmitted by forces, similarly information associations detect the resulting changes of states}. The assumption here is that the manner in which information flow occurs may be detectably different between damaged and undamaged bridges. In particular, the MI  between measurements at spatially nearby sensors pairs are dynamically different, considering a healthy bridge versus a damaged bridge. Rather than just MI of sensors with respect to each other sensors, we consider spatial pairwise MI carefully conditioning to isolate effects as we go. It is important to note that there is often a difference between direct and indirect influences.  Most notably, if there is an indirect influence, the path of information flow may have many multiple channels through the structure, and therefore, even if damage diminishes one or many of the channels, then there still may be significant information flow when an influence is indirect. Thus the change might be hard to detect or masked if simply using the pairwise MI without carefully conditioning. Therefore, it is important to identify the direct influences as the direct MI channels, as these are more sensitive to specific damage states.
By our algorithm called oMII described in Section II C that selects primary (direct) transmission channels, we consider the likely direct information coincidence of oMII as a more direct and sensitive measure  of changes.

\subsection{Optimal Mutual Information Interaction (oMII)} \label{subsec:method}

It is important to distinguish between direct and indirect influences, because a direct change or damage site may subsequently affect many further sites downstream. So by identifying the most direct influences, we hope to both identify the specific location of changes and damage, but also this prevents what otherwise would be an overly populated map of (indirect) influences making it difficult to understand clear and significant changes. The concept behind oMII is to judiciously select, based on conditional maximization of MI, a smallest number of channels that yield the largest MI, analogous to the oCSE principle that we previously developed for causality inference~\cite{sun2015causal}.
  
Therefore, in this way, the direct vibration transmission network routes in the bridge and between the sensor locations, can be identified, and most importantly, distinguished from the indirect influences. This iterative discovery process has been previously developed to identify direct causal networks from time series data~\cite{sun2015causal}.

Consider a multivariate time series $\{x_{t}^{(i)} \}$ that encodes the temporal variation of $N$ components in a system, $i=1,2,........N$. In the case of the bridge experiment, $x_{t}^{(i)}$ represents the $t$-th data point measured at the $i$-th component where the component is either the lateral or vertical acceleration collected from an individual accelerometer sensor.

For a given sensor component $i$, the oMII approach infers a set of sensor components that directly influence $i$ as follows. First, in the ``Discovery" stage (Algorithm 1), components are added one at a time to maximally reduce additional uncertainty as measured by conditional entropies, until no further reduction is possible. 
Then, in the ``Removal" stage (Algorithm 2), each component inferred from the Discovery stage is examined and removed if such removal does not result in an increase of uncertainty regarding the time variability of $i$. In both stages, a shuffle test (Algorithm 3) is used to determine whether uncertainty reduction as measured by conditional MI is statistically significant. For the results shown in this paper, we set the parameters $\theta= 0.1 $ and $N_s= 100 $ in the shuffle test.

\begin{algorithm}[H]
  \caption{:Discovery stage}
    \label{oMII algo}
  \begin{algorithmic}[1] 
  \Statex \textbf{Input:} {time series ${X_t=}\{x_t^{(i)}\}_{i=1,\dots,N;t=1,\dots,T}$ and component $i$}
    \Statex \textbf{Output:} $K_{i}$
  \State Initialize:  $K_{i} \gets \{ \emptyset \}$, $p \gets \phi$, $x \gets 1$
  \While {$x > 0$}
  \State $p \gets \argmax_{j \neq \{i, K_{i} \}} {I(X_{t}^{(i)};X_{t}^{(j)} | X_{t}^{(K_{i})}  )}$
  \If {$({X_t^{(i)};X_t^{(p)};X_t^{(K_i)}})$ passes the Shuffle Test (Algorithm 3)} 
  \State $K_{i} \gets K_{i} \cup \{ p \}$
  \Else
  \State $x\gets 0$	
		\EndIf

  \EndWhile
  \end{algorithmic}
\end{algorithm}

\begin{algorithm}[H]
  \caption{:Removal stage}
    \label{oMII algo2}
  \begin{algorithmic}[1] 
  \Statex \textbf{Input:} {time series ${X_t=}\{x_t^{(i)}\}_{i=1,\dots,N;t=1,\dots,T}$, component $i$, and set $K_i$}
    \Statex \textbf{Output:} $\hat{K_{i}}$

		\For{every $j \in K_{i}$} 
		\If {$({X_t^{(i)};X_t^{(j)};X_t^{(K_i/\{j\})}})$ fails the Shuffle Test (Algorithm 3)}
		\State $K_{i} \gets K_{i}/\{j\}$

		\EndIf
		\EndFor

		\State $\hat{K_{i}} \gets K_{i}$
  \end{algorithmic}
\end{algorithm}

\begin{algorithm}[H]
  \caption{:Shuffle test}
    \label{shuffle}
  \begin{algorithmic}[1]
    \Statex \textbf{Input:} time series ${(X^{(i)}_t=\{x_t^{(i)}\};X^{(j)}_t=\{x_t^{(j)}\};X^{(K)}_t=\{x_t^{(K)}\},~{t=1,\dots,T})}$, threshold $\theta$ and number of shuffles $N_s$
     
    \Statex \textbf{Output:} pass / fail
      \For{$\ell= 1,...,N_{s} $}
      \State generate a random permutation: $\sigma:\{1,\dots,T\}\rightarrow\{1,\dots,T\}$
      \State use $\sigma$ to obtain a shuffled time series, ${Y_t=}\{y_t\}$, where $y_t\gets x^{(j)}_{\sigma(t)}$
		 	\State compute $I_{\ell} \gets I({X_{t}^{(i)}, Y_t | X_t^{(K)}})$
       \EndFor
       \State $S\gets\mbox{the~}\lfloor(1-\theta)N_s\rfloor\mbox{th largest value from~}\{I_1,\dots,I_{N_s}\}$
       \If{$I({X_t^{(i)};X_t^{(j)}|X_t^{(K)}})>S$}
	\State \textbf{output:} pass
	\Else 
	\State \textbf{output:} fail
	\EndIf

  \end{algorithmic}
\end{algorithm}

\section{Waddington Bridge Data: Description and Basic Statistical Properties} \label{sec:damage}

\subsection{Description of the Waddington Bridge Data}
In this section we describe the bridge that was used for damage detection, the instrumentation setup, and the levels of damage introduced to the structure. 

\subsubsection{The Waddington Bridge}
\begin{figure}[htbp]
\centering
\includegraphics[width=0.6\textwidth]{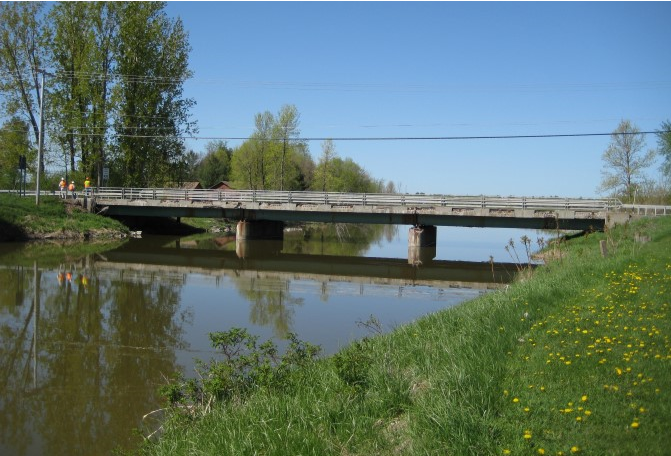} 
\caption{The Waddington Bridge, in New York State Route 345 over Big Sucker Brook in the town of Waddington, NY.}
\label{fig:Bridge}
 \end{figure}

The Waddingtonbridge, constructed in 1957, is located in New York State Route 345 over Big Sucker Brook in the town of Waddington, NY (see Fig.~\ref{fig:Bridge}). The highway bridge investigated consists of a 19.1cm (7.5 in.) thick reinforced concrete slab supported by three interior $ W33\times152$ and exterior $W33\times141$ steel girders over each span. The bridge has two-lane structures consisting of three 13.7 m (45 ft) simply supported spans carrying a total span of 41.7 m (137 ft) at an elevation of approximately 1.2 m (4 ft) from waterline. The girders have a center-to-center spacing of 2.1m (7 ft) and are supported by fixed and rocker steel bearings.

\subsubsection{Instrumentation}
In this case study, we use wireless sensor solution (WSS) for bridge health monitoring and condition assessment.  The WSS was developed at Clarkson University as a versatile wireless sensing platform optimized for large scale, high rate, real-time acquisition \cite{kim2007health,harms2010structural}. The system was developed based on off-the-shelf components to provide a low-power sensing interface for vibration, strain, and temperature measurements with signal conditioning tailored to the typical highway bridge response spectrum. The wireless communication is facilitated by a low-power chip transceiver employing direct sequence spread spectrum modulation over a 2.4GHz carrier frequency. 
Proprietary embedded software was used to sample data at an effective rate 128Hz.
          
The bridge span was instrumented in a rectangular grid array at 30 locations with dual-axis (vertical and lateral) accelerometers, in effect resulting in 60 vibration sensors.

\begin{figure}[htbp]
\centering
\subfigure[~{(Top view) Physical spatial layout of the indexed accelerometers, within the bridge structure. The bridge is divided into three sections. When a truck goes by, its goes through all three sections. In this particular experiment, the sensors are placed to cover one of these sections near one end of the bridge.}]{\includegraphics[width=8cm]{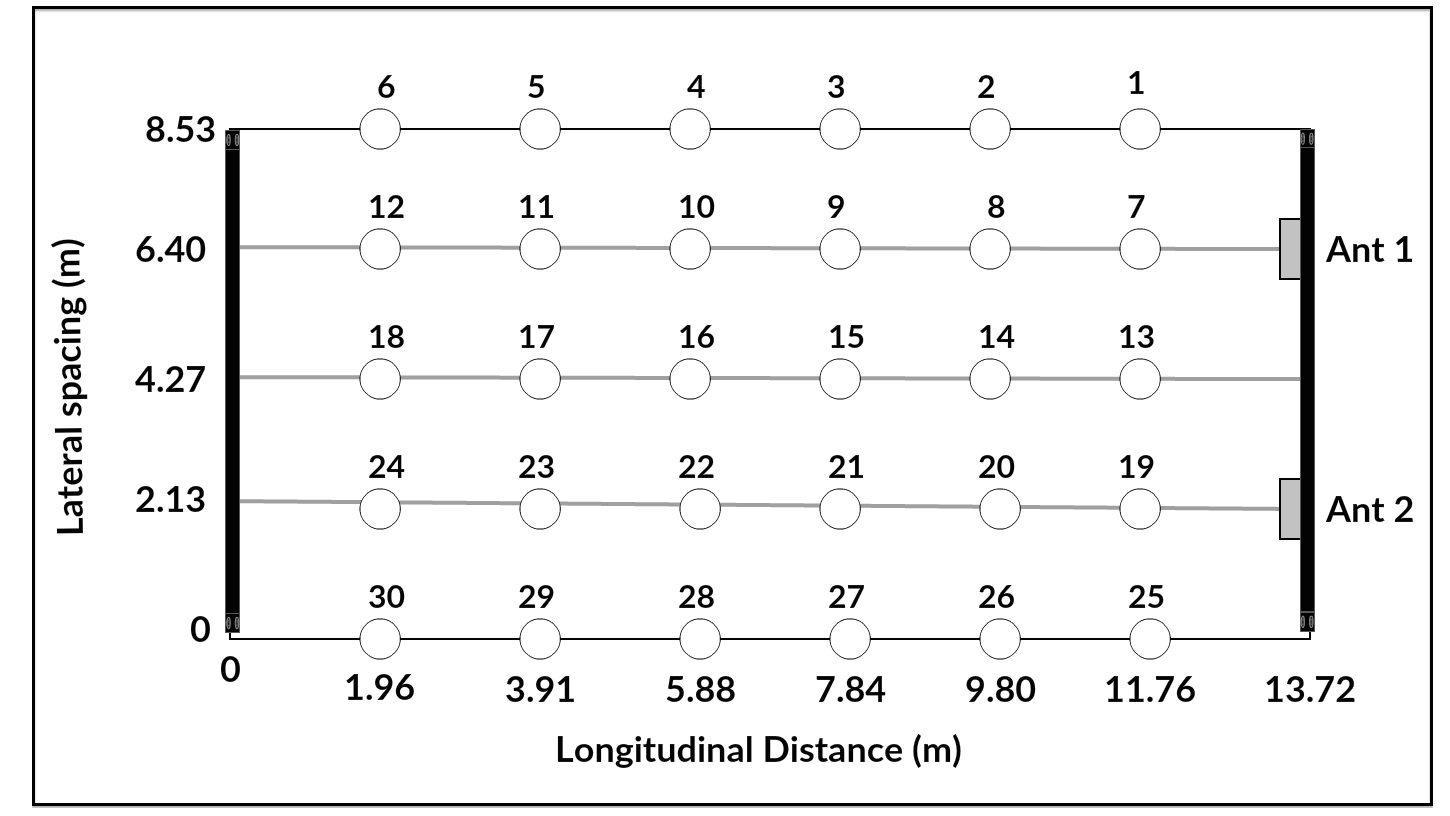}}
\subfigure[~{Top view with the same orientation and coverage area as in (a), here showing the location of the 1st diaphragm connections where the damages are introduced in the experiment.}]{\includegraphics[width=8.1cm]{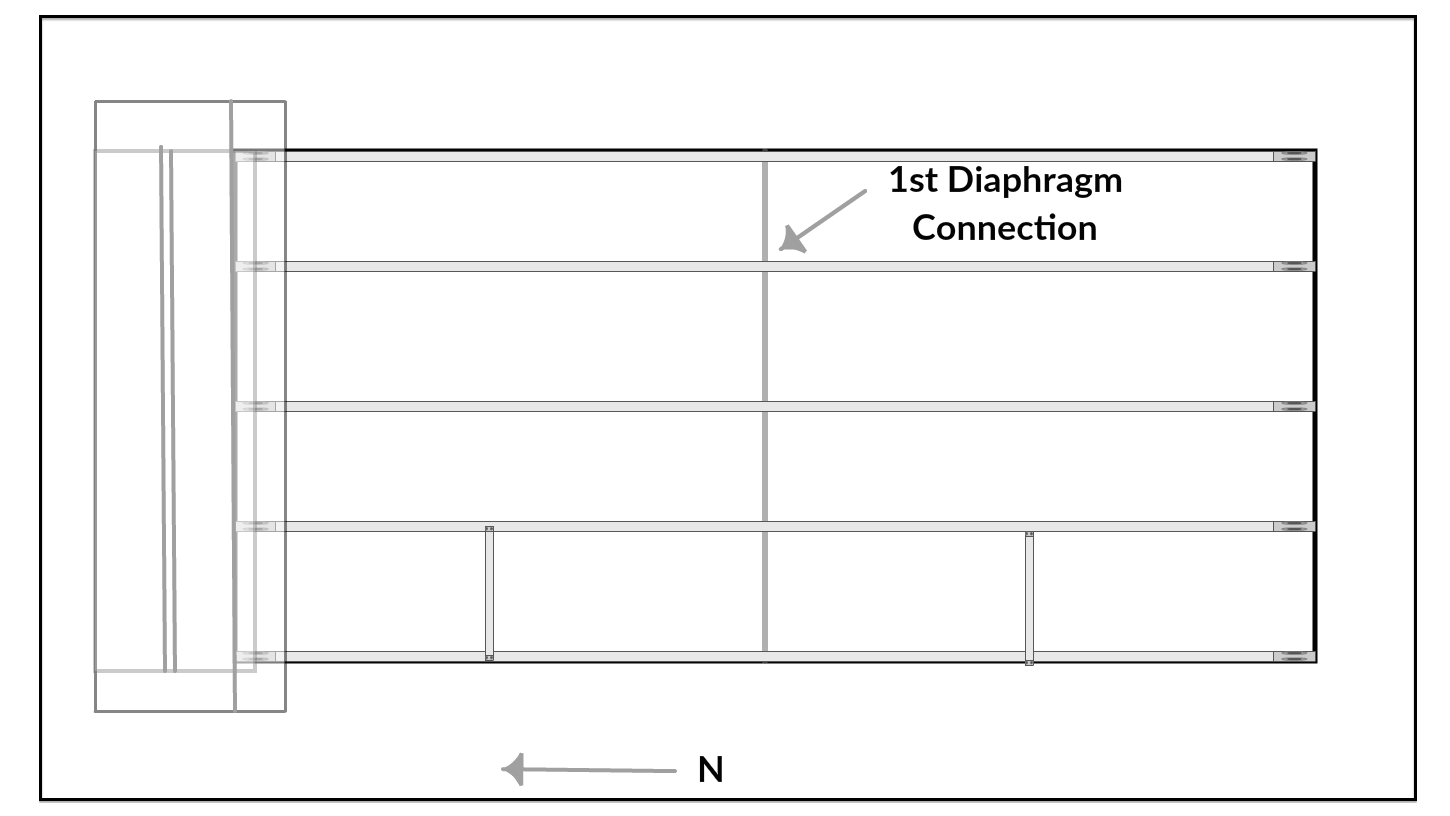}}
\subfigure[~Vibrational energy is introduced to the bridge in a controlled manner by driving a truck over the structure, both before and after damage has been introduced]{\includegraphics[width=8cm]{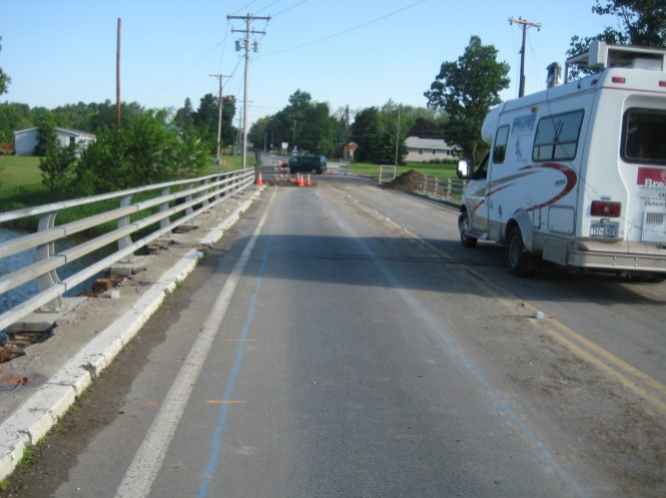}}
\caption{Physical layout of the accelerometers and the field test vibration introduction.}
\label{damage_intro}
\end{figure}

The sensor locations are shown in the Fig.~\ref{damage_intro}(a). The lateral and longitudinal spacing between these accelerometers are 2.13m and 1.96m, respectively.

\vspace{1cm}

\subsubsection{Field Testing and Damage Introduction} \label{field-test}
The sequence of tests performed is outlined in Table 1. Each test consisted of the acquisition of approximately 90-second time history. Each case included three passes across the bridge with a truck in both directions. More explicitly, first test was performed for $\sim 90$ seconds, and measurements of the sensors were taken after first pass of the truck, the second set of test measurements was taken after the second pass of the truck for another $\sim 90$ seconds, and so on. In total 9 tests were performed.

Peak acceleration induced by the truck loading, as measured across the sampled locations, was generally 15mg, while peak lateral acceleration of 7mg was typical. The damage test was done with 6 bolts in 1st diaphragm connections (see Fig.~\ref{damage_intro}(b)).

\begin{table}[H]
\begin{center}
 \begin{tabular}{||c | c | c | c||} 
 \hline
 Case & Test ID & Scenario & Comments \\ [0.5ex] 
 \hline\hline
 1 & 1-3& Baseline & ``Healthy'' Structure \\ 
 \hline
 2& 4-6 & 1st diaphragm & Removal of four out of six bolts (Damage 1)\\
 \hline
 3 & 7-9 & 1st diaphragm & Removal of all six out of six bolts (Damage 2) \\
 \hline
\end{tabular}
\end{center}
\caption{Baseline and Damage Test Scenarios.}
\label{tests}
\end{table}

For each test ($\sim90$ seconds), lateral and vertical accelerations are measured and recorded at 128HZ, producing a raw time series $\{\tilde{x}^{(k)}_t\}$ for each sensor, where $k$ denotes the index of the sensor as labeled in Fig.~\ref{damage_intro}(a) and $t=1,2,\dots,T=11536$.

We standardize each raw time series by a linear transformation
\begin{equation}
	x^{(k)}_t=\left(\tilde{x}^{(k)}_t-\mu(\tilde{x}^{(k)})\right)/\sigma\left(\tilde{x}^{(k)}\right),
\end{equation}
where $\mu(\cdot)$ and $\sigma(\cdot)$ denote the empirical mean and standard deviation of the given time series $\{x_t\}_{t=1}^{T}$, that is, 
$\mu(x)=\frac{1}{T}\sum_{t=1}^{T}x_t$ and $\sigma(x)=\sqrt{\frac{1}{T-1}\sum_{t=1}^{T}(x_t-\mu(x))^2}$.
Such transformation produces time series that have zero mean and unit variance. For the remainder of the paper we shall always deal with such standardized time series $\{x^{(k)}_t\}$.

\subsection{Basic Statistical Findings - Laplace Distribution} \label{subsec: Laplace}

\begin{figure}[htbp]
\centering
\subfigure[~Baseline - Lateral direction]{\includegraphics[width=0.35\textwidth]{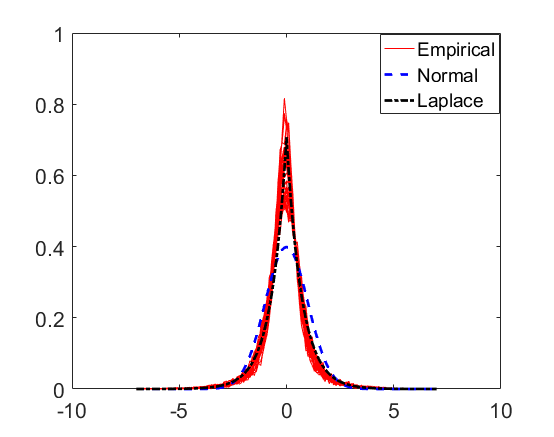}}
\subfigure[~Baseline - Vertical direction]{\includegraphics[width=0.35\textwidth]{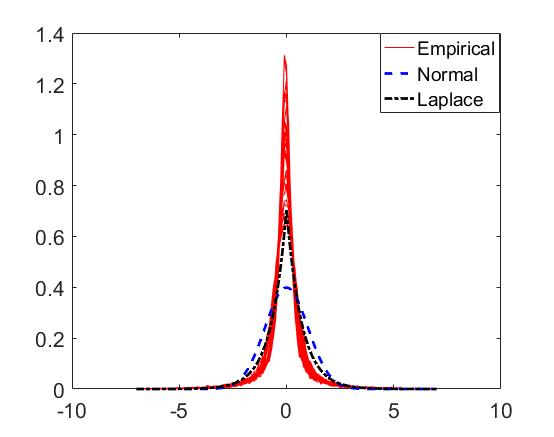}}
\subfigure[~Damage 1 - Lateral direction]{\includegraphics[width=0.35\textwidth]{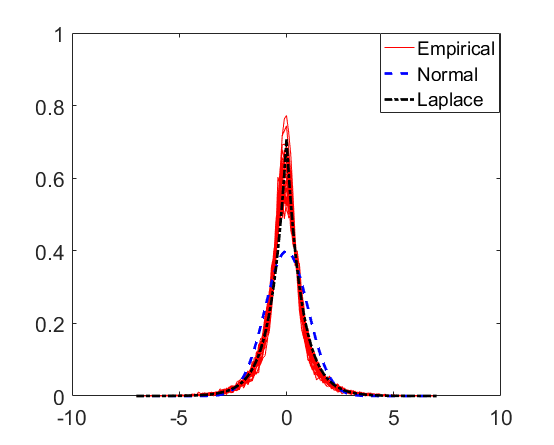}}
\subfigure[~Damage 1 - Vertical direction]{\includegraphics[width=0.35\textwidth]{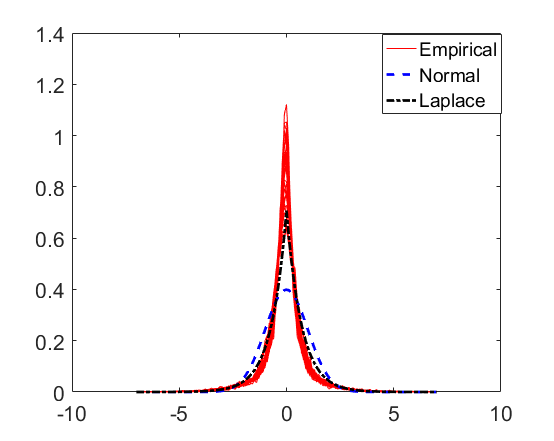}}
\subfigure[~Damage 2 - Lateral direction]{\includegraphics[width=0.35\textwidth]{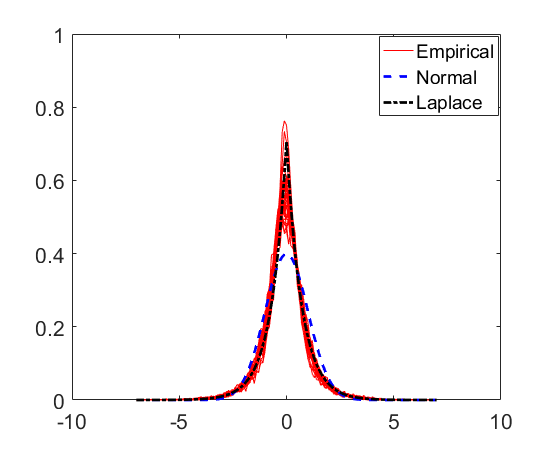}}
\subfigure[~Damage 2 - Vertical direction]{\includegraphics[width=0.35\textwidth]{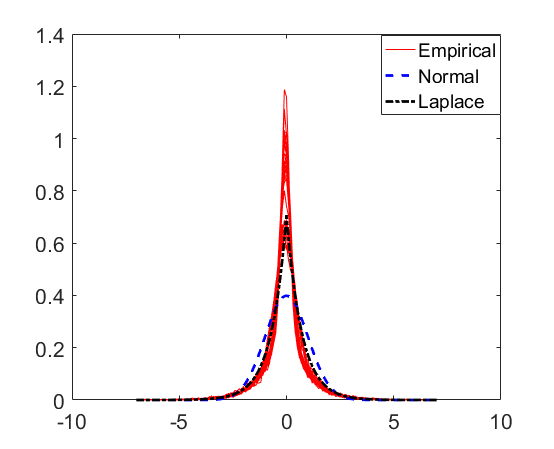}}
\caption{Probability distributions of the sensor time series after first pass of the truck. (a) Baseline-lateral, (b) Baseline-vertical, (c) Damage 1-lateral, (d) Damage 1-vertical, (e) Damage 2-lateral, (f) Damage 2-vertical. In all panels we also plot the normal distribution and the Laplace distribution for visual comparison. All distributions have been standardized to have zero mean and unit variance.}
\label{fig:distribution1}
\end{figure}

Since the information theoretic entropies are in terms of probabilities, there is the necessity of ``good'' statistical estimation from sampled time series data. Recall that there is a total of 30 sensors, and 2 time series are measured and recorded at each sensor (accelerometer) location: lateral acceleration and vertical acceleration.
Fig.~\ref{fig:distribution1} shows the distribution of the sensor time series plotted against two baseline distributions: the normal distribution and the Laplace distribution, both standardized to have zero mean and unit variance, given below:

\begin{equation}\label{eq:normalLaplace}
\begin{cases}
\mbox{(Normal distribution)~}f(x|\mu,\sigma^2) &= \frac{1}{\sigma \sqrt{2\pi}}\exp{\frac{(x-\mu)^2}{2 \sigma^2}},\\
\mbox{(Laplace distribution)~}f(x|\mu,b) &= \frac{1}{2b}\exp{\frac{-|x-\mu|}{b}},
\end{cases}
\end{equation}
with mean $\mu=0$, $\sigma=1$ (normal distribution), and $b=\sqrt{2}/2$ (Laplace distribution).

\begin{figure}[htbp]
\centering
\subfigure[~Baseline - Lateral direction]{\includegraphics[width=0.4\textwidth]{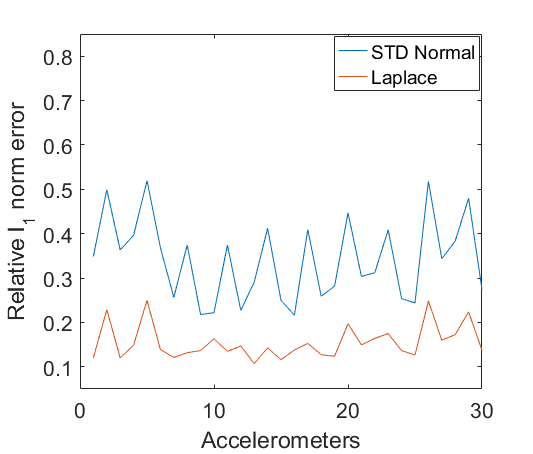}}
\subfigure[~Baseline - Vertical direction]{\includegraphics[width=0.4\textwidth]{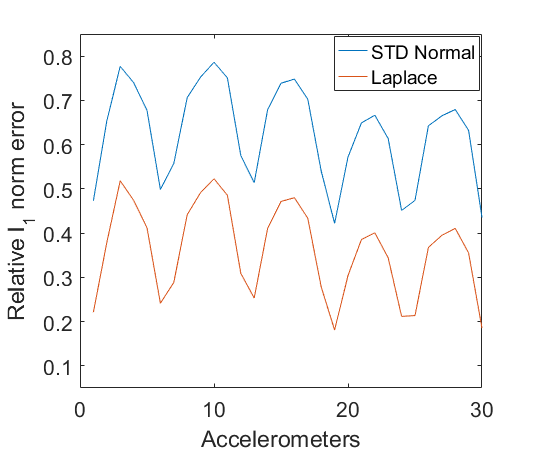}}
\subfigure[~Damage 1 - Lateral direction]{\includegraphics[width=0.4\textwidth]{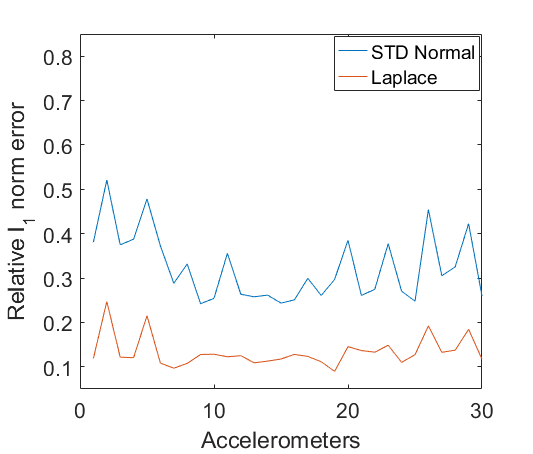}}
\subfigure[~Damage 1 - Vertical direction]{\includegraphics[width=0.4\textwidth]{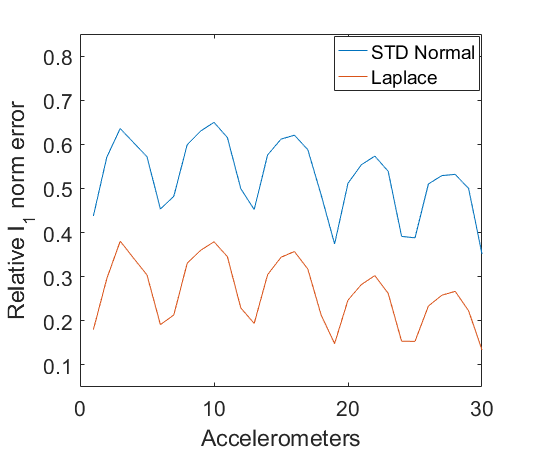}}
\subfigure[~Damage 2 - Lateral direction]{\includegraphics[width=0.4\textwidth]{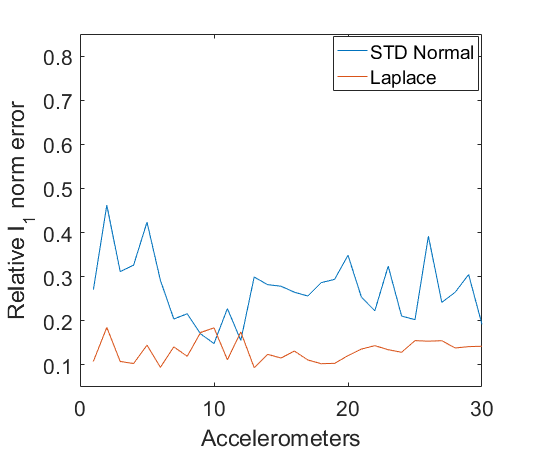}}
\subfigure[~Damage 2 - Vertical direction]{\includegraphics[width=0.4\textwidth]{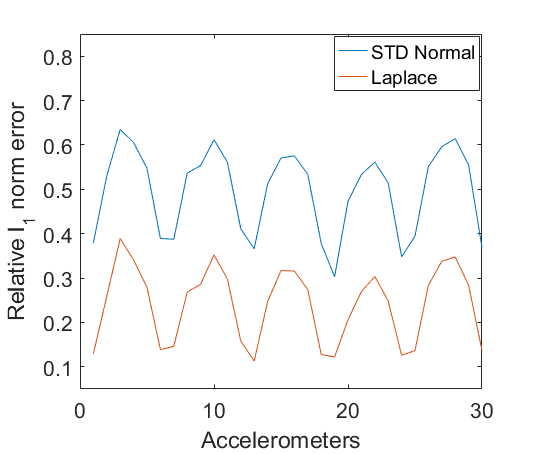}}
\caption{Relative $l_{1}$-norm error of the probability distributions of the sensor time series after first pass of the truck. Here relative $l_{1}$-norm error is defined by $|| p_{1} - p_{2}|| / || p_{1} || $, where $p_1$ is the empirical distribution and $p_2$ is either standard normal or Laplace distribution. (a) Baseline-lateral, (b) Baseline-vertical, (c) Damage 1-lateral, (d) Damage 1-vertical, (e) Damage 2-lateral, (f) Damage 2-vertical. In all panels the distribution error is computed against both the standard normal and the Laplace distribution.}
\label{fig:distribution_error}
\end{figure}

From the figure it is visually evident that the measured acceleration data more closely follow Laplace distribution than normal distribution. To draw this conclusion from a quantitative standpoint, we compute the $l_1$ norm between the standardized distribution of the time series for each sensor component and either a Laplace distribution or a normal distribution. The results are shown in Fig.~\ref{fig:distribution_error}, confirming that Laplace distribution is generally a better fit for the observed data across all sensors in all test scenarios.

\section{Results} \label{sec:results}
\subsection{Parametric Entropy Estimator for Multivariate Laplace Distribution}
In section III B, we established that the lateral and vertical accelerations recorded by the bridge sensors more closely follow Laplace distribution than normal distribution. (Assuming ) a parametric estimator for entropy generally requires fewer data points for the same level accuracy.

If the underlying data follows a normal distribution, there is a parameter estimator based on a closed-form formula that only involves the variance (and covariance for multivariate data) of the time series. However, no such closed-form formula exists for the Laplace distribution. Our strategy is to use a Monte Carlo method to numerically evaluate the integrals which define entropies and mutual information, where the pdf in the integrals are assumed to be Laplace distributions whose covariances are directly estimated from data. Since there are multiple variables involved, the assumed distribution is a multivariate Laplace distribution whose pdf is given by~\cite{eltoft2006multivariate}
\begin{align}
f_{X} \left(\textbf{x} \right) = \frac{1}{{2\pi}^{(d/2)}} \frac{2}{\lambda} \frac{K_{(d/2) - 1} \left( \sqrt{\frac{2}{\lambda} q(\textbf{x})} \right)}{\left( \sqrt{\frac{\lambda}{2} q(\textbf{x})} \right)^{(d/2) - 1}},
\end{align}
where $\textbf{x} \in\mathbb{R}^d$, $K_{(d/2)-1}$ is the modified Bessel function of the second kind with order $(d/2)-1$ evaluated at $\textbf{x}$,
\begin{equation}
q(\textbf{x}) = \lambda\left(\textbf{x} - \mu \right)^{\top} \Sigma^{-1} \left(\textbf{x} - \mu\right),
\end{equation}
$\mu$ is the mean vector, $\Sigma$ is the covariance matrix and $\lambda=\sqrt{\mbox{det}(\Sigma)}$.

\subsection{Spatial Pairwise MI - Baseline, Damage 1, Damage 2}
\begin{figure}[htbp]
 \centering
 \includegraphics[width=0.8\textwidth]{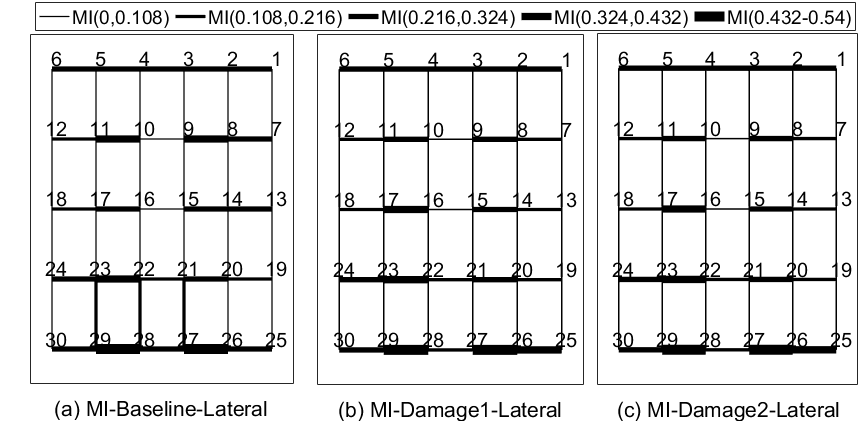} 
\caption{Spatial pairwise mutual information  between accelerometers in the lateral direction after first pass of the truck. Fig. (a) healthy bridge (b) Damaged bridge after removing 4 bolts of six bolts (c) Damaged bridge after removing all 6 bolts from the first diaphragm. There is a significant loss of mutual information in both damaged bridges compares with healthy bridge.}
\label{fig:PMI}
\end{figure}

\begin{figure}[htbp]
 \centering
 \includegraphics[width=0.8\textwidth]{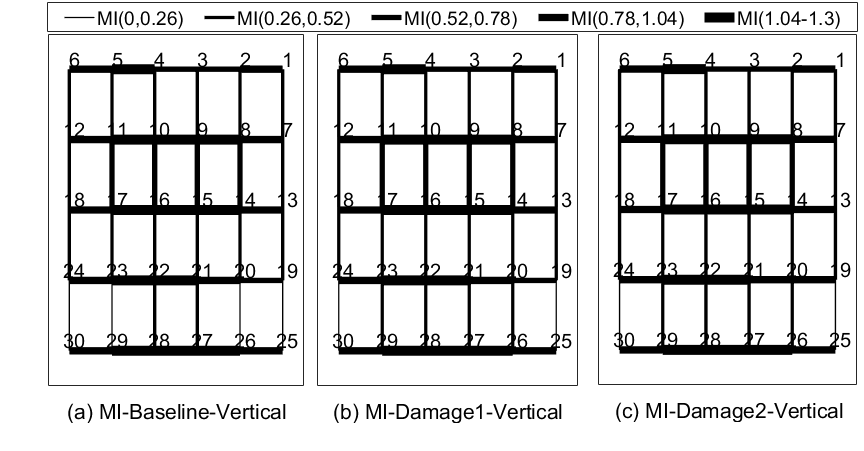} 
 \caption{Spatial pairwise mutual information  between accelerometers in the vertical direction after first pass of the truck. Fig. (a) healthy bridge (b) damaged bridge after removing 4 bolts of six bolts  (c) damaged bridge after removing all 6 bolts from the first diaphragm. }
\label{fig:PMIV}
 \end{figure}

\begin{figure}[htbp]
  \centering
  \includegraphics[width=0.8\textwidth]{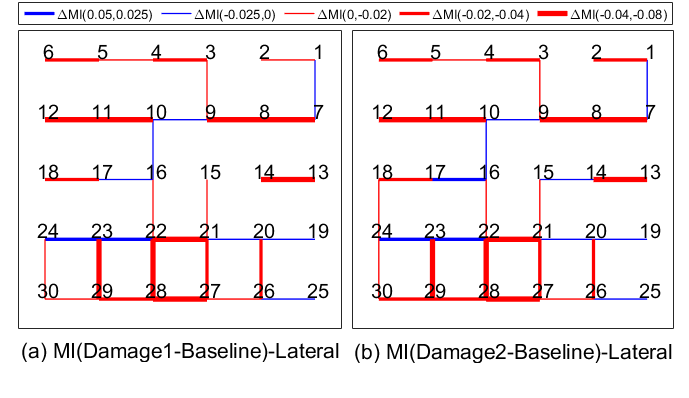} 
 \caption{Difference of the spatial pairwise mutual information between accelerometers in the lateral direction after 1st pass of the truck. (a) difference of  healthy bridge and bridge after first damage (b) difference of  healthy bridge and bridge after second damage. There is some difference of connection strengths after first damage and second damage.}
\label{fig:MIDIFL1}
 \end{figure}

This section illustrates use of pairwise mutual information to study the damage detection of the bridge. 
Bridge layout has 30  accelerometers. Each accelerometer records time series data for vertical and lateral directions, each. We use the spatial pairwise mutual informations as a probe to study the difference between healthy bridge structure and damaged bridge structures. 
In particular, for each sensor, we compute the MI between the time series produced by it and those from each of its 4 spatially nearest neighbors.
All the results here are shown only for the scenarios that we call, baseline, damage 1, damage 2 after the first pass of the truck (tests 1, 4, 7 in Table~\ref{tests}).
Fig.~\ref{fig:PMI} shows the estimated pairwise mutual information of the lateral-direction accelerometer for the first pass of the truck in both directions. In particular, Fig.~\ref{fig:PMI}(a) describes the mutual interactions for baseline, which is referred as the healthy bridge. Here the thickness of the lines are drawn proportional to the corresponding pairwise mutual information. Maximum pairwise mutual information is found between accelerometers 28-29, 26-27. Interactions are less than 0.108 between accelerometer 9-10 and 15-16. Fig.~\ref{fig:PMI}(b) describes the mutual interactions for the partially damaged bridge, which is referred as the first damage of study of the bridge. Notice that, first damage study was done by removing 4 bolts (out of 6) from the first diaphragm. We can see that after the first damage, interactions remain in the same range for almost all the accelerometers. However, there is a loss of MI between 10-11 and 8-9, which can be identified clearly in Fig.~\ref{fig:MIDIFL1}. Further damage to the bridge in the tests was done by removing all 6 bolts from first diaphragm, with the resulting pairwise mutual information shown in Fig.~\ref{fig:PMI}(c). Similar plots are shown with respect to the vertical accelerations in Fig.~\ref{fig:PMIV}.

\begin{figure}[htbp]
 \centering
  \includegraphics[width=0.8\textwidth]{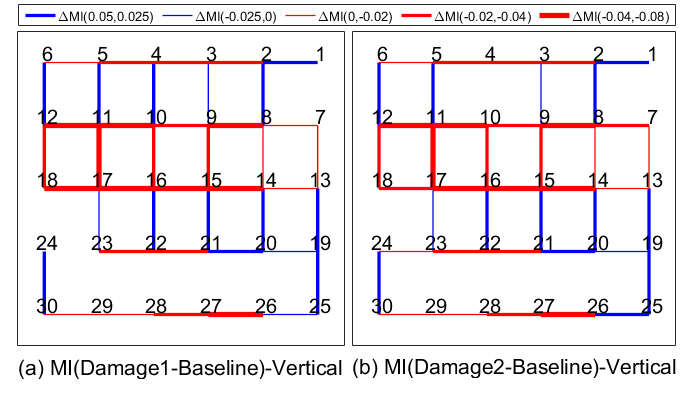} 
 \caption{ Difference of the spatial pairwise mutual information between accelerometers in the vertical direction after 1st pass of the truck. (a) difference of  healthy bridge and bridge after first damage (b) difference of  healthy bridge and bridge after second damage. There is some difference of connection strengths after first damage and second damage in the vertical direction.}
 \label{fig:MIDIFV1}
 \end{figure}

To more clearly illustrate the difference in information association between nearby spatial sites, we further plot the difference of spatial pairwise mutual information between the healthy bridge and damaged bridges, as shown in Fig.~\ref{fig:MIDIFL1} (lateral direction) and Fig.~\ref{fig:MIDIFV1} (vertical direction). In these figures, red and blue are used to respectively denote negative and positive changes due to the damage.
Several observations are in order. First, (with very few exceptions) damage to the structure (as achieved by the removal of bolts in the tests) seems to generally lower the value of mutual information in the lateral direction between spatially nearby sites, indicating a lower coupling, and such change is enhanced with further structural damage [Fig.~\ref{fig:MIDIFL1}]. 

Secondly, some difference in connection strengths  can be seen in the both Fig.~\ref{fig:MIDIFL1} and Fig.~\ref{fig:MIDIFV1}. For example in Fig.~\ref{fig:MIDIFL1}(a) 11-12, 10-11, 8-9, 7-8 and in Fig.~\ref{fig:MIDIFV1}(a) 11-12, 10-11, 8-9, 16-17.
Same mutual interaction strength differences between the above mentioned accelerometers can be seen in Fig.~\ref{fig:MIDIFL1} (b) and  Fig.~\ref{fig:MIDIFV1}(b), which describes the differences in information flow between the healthy bridge and the bridge after second damage in the lateral and vertical directions, respectively. 
However, some connections remain the same as measured by the spatially pairwise MI. Ex: see the connection 9-15 in the lateral direction [Fig.~\ref{fig:MIDIFL1}] and the connection 1-7 in the vertical direction [Fig.~\ref{fig:MIDIFV1}]. 

\subsection{oMII - Baseline, Damage 1, Damage 2}
\begin{figure}[htbp]
\centering
\subfigure[~Baseline ]{\includegraphics[width=8cm]{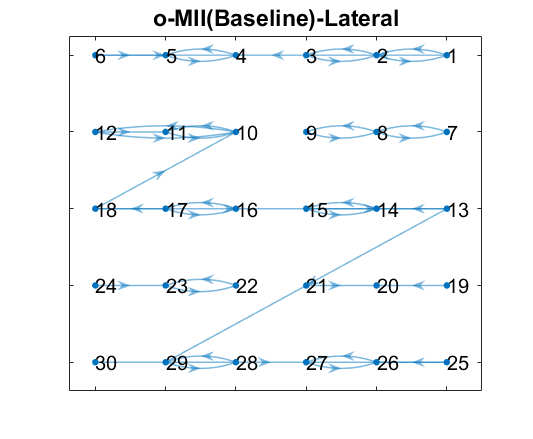}}
\subfigure[~Damage 1]{\includegraphics[width=8cm]{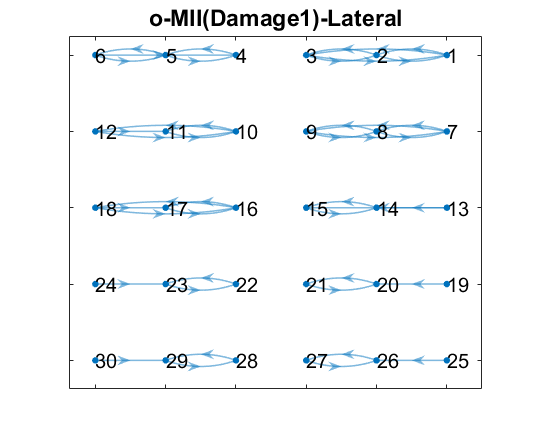}}
\subfigure[~Damage 2]{\includegraphics[width=8cm]{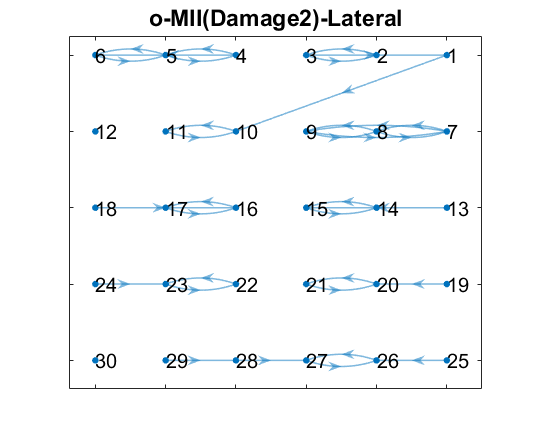}}
 \caption{ Optimal mutual information interaction between accelerometers in the lateral direction after first pass of the truck. Fig. (a) healthy bridge (b) damaged bridge after removing four bolts from 6 bolts (c) damaged bridge after removing all the 6 bolts from the first diaphragm.}
\label{fig:o-MIIL}
\end{figure}

\begin{figure}[htbp]
\centering
\subfigure[~Baseline ]{\includegraphics[width=8cm]{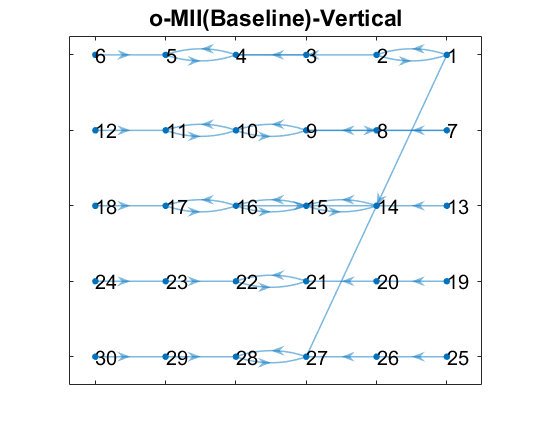}}
\subfigure[~Damage 1]{\includegraphics[width=8cm]{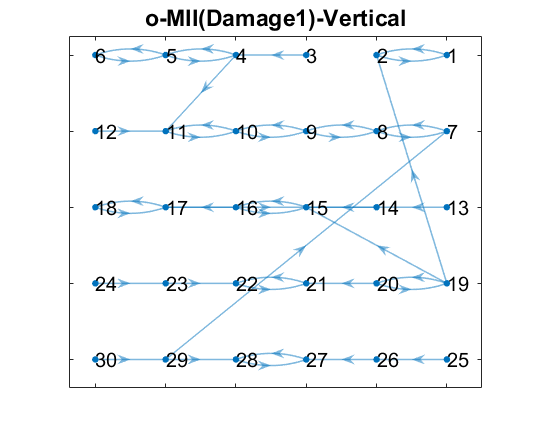}}
\subfigure[~Damage 2]{\includegraphics[width=8cm]{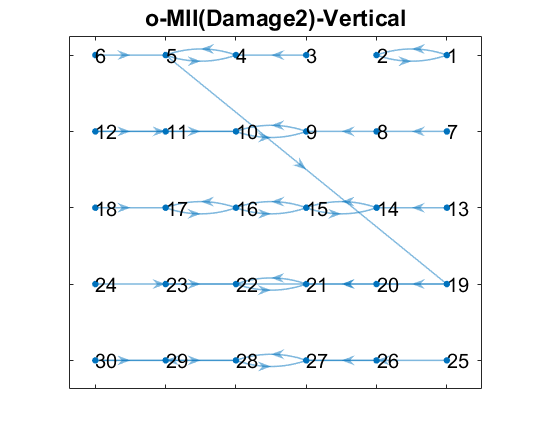}}
 \caption{ Optimal mutual information interaction between accelerometers in the vertical direction after first pass of the truck. Fig.(a) healthy bridge (b) damaged bridge after removing four bolts from 6 bolts (c) damaged bridge after removing all the 6 bolts from the first diaphragm.}
\label{fig:o-MIIV}
\end{figure}

In section IV B we use spatial pairwise MI to study damage detection of the bridge structure. However, pairwise mutual information itself cannot differentiate between direct and indirect couplings or select the most influential pairs among all possible couplings. The oMII connections by contrast   identify direct influences, whether they be spatially nearest neighbors or more remote sites on the bridge structure. It is particularly valuable in a damage detection scenario where  oMII  connections may be especially fragile to changes in the structure.
Applying the oMII algorithm as described in Section~\ref{subsec:method} to the lateral time series with parameters  $\theta = 0.1$ for the stopping criteria and $N_s=100$ in the shuffle test, we show the resulting direct connections in Fig.~\ref{fig:o-MIIL} and Fig.~\ref{fig:o-MIIV} for lateral and vertical directions. 

Figure~\ref{fig:o-MIIL} (a) shows the optimal mutual information interactions for the healthy bridge. It can be seen from the figure that the bridge structure supports more information flow in the same direction as the truck lanes (horizontal direction in the figure). Exceptions are near the center of the bridge which could have been due to the first diaphragm placed in the bridge structure (see Fig.~\ref{damage_intro}(b)). The optimal mutual information interactions after the first damage (removal of 4 out of 6 bolts) and second damage (removal of all 6 bolts) are shown in Figs.~\ref{fig:o-MIIL}(b) and ~\ref{fig:o-MIIL}(c).. After the first  damage, we can see there is both a loss of oMII connections and new ones. This indicates that the loosening of the bridge potentially allows for new pathways for vibrational signals to propagate. Clear change in the information flow in the center of the bridge can be seen from the Fig.~\ref{fig:o-MIIV}. However, in the vertical axis also bridge structure supports more information flow in the same direction as the truck lanes. 

\begin{figure}[htbp]
\centering
\subfigure[~Baseline - Damage 1 ]{\includegraphics[width=8cm]{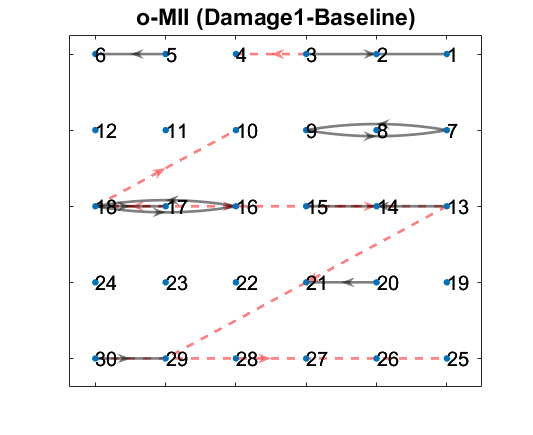}}
\subfigure[~Baseline - Damage 2]{\includegraphics[width=8cm]{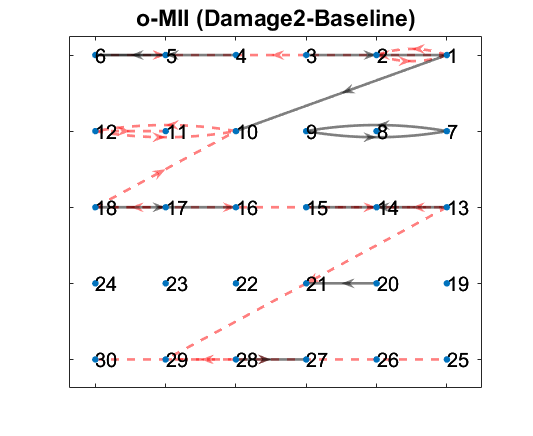}}
 \caption{ Difference of the oMII between baseline and damaged bridges in the lateral direction after 1st pass of the truck. (a) difference of  healthy bridge and bridge after first damage (b) difference of  healthy bridge and bridge after second damage. Red and black lines represent new connections and loss connections after damaged the bridge respectively.}
\label{fig:MIDIFL}
 \end{figure}

\begin{figure}[htbp]
\centering
\subfigure[~Baseline - Damage 1 ]{\includegraphics[width=8cm]{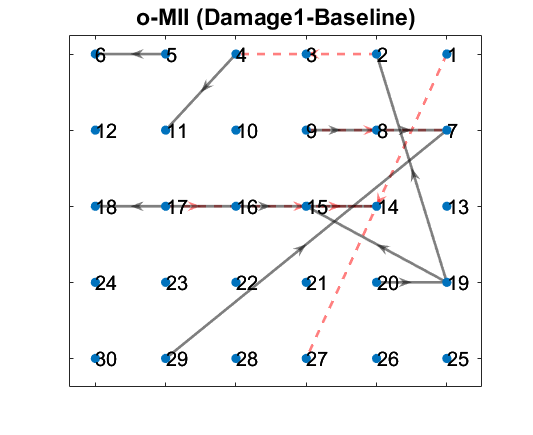}}
\subfigure[~Baseline - Damage 2]{\includegraphics[width=8cm]{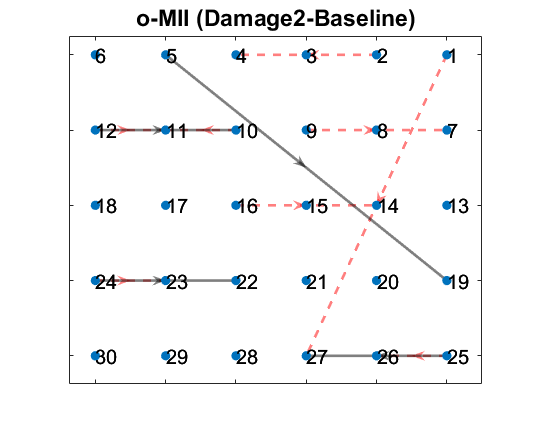}}
 \caption{ Difference of the oMII between baseline and damaged bridges in the vertical direction after 1st pass of the truck. (a) difference of  healthy bridge and bridge after first damage (b) difference of  healthy bridge and bridge after second damage. Red and black lines represent new connections and loss connections after damaged the bridge respectively.}
\label{fig:MIDIFV}
 \end{figure}

To more clearly see which oMII connections are lost/created after damage, we compute and plot the difference of the oMII connections between the healthy bridge and the damaged bridges in Fig.~\ref{fig:MIDIFL}(a) (healthy vs. first damage) in lateral direction, Fig.~\ref{fig:MIDIFL}(b) (healthy vs. second damage) in lateral direction, \ref{fig:MIDIFV}(a) (healthy vs. first damage) in vertical direction, and  Fig.~\ref{fig:MIDIFV}(b) (healthy vs. second damage) in vertical direction, respectively. In these figures, the lost connections are shown as dashed red lines and the new ones are illustrated by solid black lines. Directionality is denoted by arrows. One can see that there are 10 new connections (solid black lines) and 7 loss connections (dashed red lines) appear after the first damage in Fig.~\ref{fig:MIDIFL}(a).

 Due to the second damage, we can see there are significant changes in the information transfer between sensors:10 new connections are occurred and 16 connections are lost after the second damage (see \ref{fig:MIDIFL}(b)). We can see from the \ref{fig:MIDIFV}(a) and \ref{fig:MIDIFV}(b) that the number of changes of connections in the vertical direction and lateral direction remain same after the first damage. After the second damage, number of new and loss connections ( so number of changes) have decreased in the vertical direction. However, the number of changes in the connections have increased in the lateral direction.  
 
The results obtained by oMII as plotted in Fig.~\ref{fig:MIDIFL} and Fig.~\ref{fig:MIDIFV} show that for the vertical ``gap" in the middle of the region which corresponds to where the diaphragm connections are, new connections are formed after damages are introduced. Interestingly, this pattern only appears for the lateral accelerations but not the vertical ones.
 
\begin{figure}[htbp]
\centering
\subfigure[~Degree distribution - Lateral ]{\includegraphics[width=8cm]{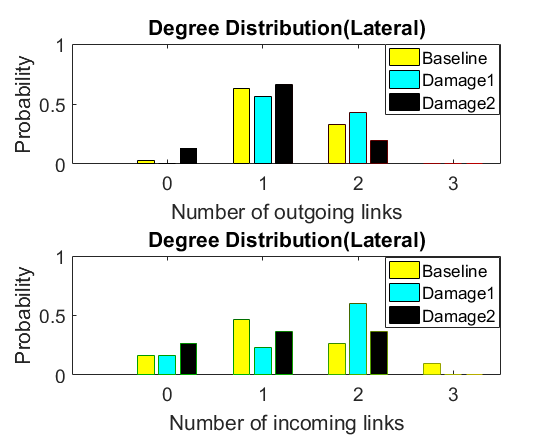}}
\subfigure[~Degree distribution - Vertical ]{\includegraphics[width=8cm]{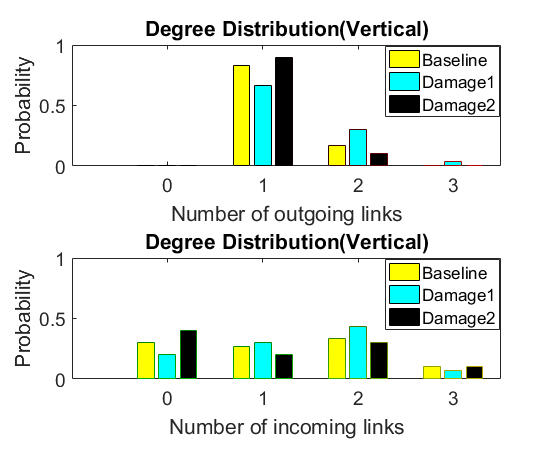}}
 \caption{ Degree distribution of the oMII based networks of the baseline and damaged bridges after 1st pass of the truck in the (a) lateral direction (b) vertical direction. 0, 1, 2, 3 are the number of outgoing and incoming links. }
\label{fig:degree}
 \end{figure}

Another way to characterize the results is to look at degree distribution of outgoing and incoming links. The degree distributions in the baseline and damaged bridge are shown in the Fig.~\ref{fig:degree}. Yellow, cyan, and black bars represent those corresponding to the baseline, damage 1, damage 2 respectively. For some scenarios, there are some sensors that do not have outgoing or incoming links. The maximum number of outgoing and incoming link is 3 for all structures. 
In both the lateral and vertical directions, the probability of having no incoming links and the probability of having no outgoing links have increased after damage 2 (as compared to the healthy bridge). This implies that after introducing relatively large damage to the bridge there is a significant number of locations that become ``disconnected" from the rest of the bridge in terms of information flow.

 
In summary, we observe that there are both vanishing and new direct connections between sensors as the damage experiment is progressed (Fig.~\ref{fig:degree}), inferred as direct connections of the accelerometer signals by vibrational transmission of energy, and furthermore the sensed values are relatively stable across the repeated experiment.  Therefore, we have described a simple experimental protocol, driving a truck over a bridge with instrumentation, and a information theoretic approach to interpret the data, that is positively indicative of potential structural damages.

\section{Discussion and Conclusion}
In this work, we used an MI based approach to study damage detection of a bridge located in Waddington, New York. The damage to the bridge was introduced by removing bolts from the first diaphragm of the bridge and a sequence of tests were performed with time series data collected on various locations on the bridge. In particular, comparing to the baseline case where no damage was introduced, two levels of damage were tested by either removing four out of the six bolts (damage 1) or all six bolts (damage 2).

Our first finding is that the measured data, which are accelerations detected by sensors on the bridge, more closely follow Laplace distribution than normal distribution. This enables us to develop parametric estimators of mutual information and conditional mutual information that are more efficient than non-parametric ones.

Our second finding is that the spatial nearest-neighbor interactions as measured by mutual information tend to become weaker as more damage is imposed. This is consistent with the intuition that less force and energy pass between adjacent sites as the bridge is ``loosened" due to the removal of bolts.

Finally, we found that the more primary direction of direct influence and information flow as detected by oMII goes in the direction of traffic flow even after partial damage to the bridge. Based on the particular experiments from which these results are obtained, it is not yet clear whether such unidirectional dominance of information flow comes from the underlying mechanical structure or from the effect of drive-through by the trucks.

Among the many unsolved problems, we note that it is important to design experiments for which results from noninvasive damage detection techniques such as the ones investigated herein can be experimentally validated. The success of such validation is necessary for making reliable assessment of the structural fatigue as well as the risk of sudden and disastrous collapse of bridges.

\section*{Acknowledgments}

This work has been supported by  the Army Research office under Grant No. W911NF-12-1-0276 and No. W911NF-16-1-0081, and the Simons Foundation under Grant No. 318812.

The authors thank Matthew Whelan and Michael Gangone for the instrumentation and field testing in the bridge.

\bibliographystyle{ieeetr}

\end{document}